\documentstyle[psfig,twoside]{article}

\catcode`\@=11
\long\def\@makefntext#1{
\protect\noindent \hbox to 3.2pt {\hskip-.9pt  
$^{{\eightrm\@thefnmark}}$\hfil}#1\hfill}               

\def\@makefnmark{\hbox to 0pt{$^{\@thefnmark}$\hss}}    
        
\def\ps@myheadings{\let\@mkboth\@gobbletwo
\def\@oddhead{\hbox{}
\rightmark\hfil\eightrm\thepage}   
\def\@oddfoot{}\def\@evenhead{\eightrm\thepage\hfil
\leftmark\hbox{}}\def\@evenfoot{}
\def\sectionmark##1{}\def\subsectionmark##1{}}

\oddsidemargin=\evensidemargin
\newcounter{sectionc}\newcounter{subsectionc}\newcounter{subsubsectionc}
\renewcommand{\section}[1] {\vspace{12pt}\addtocounter{sectionc}{1} 
\setcounter{subsectionc}{0}\setcounter{subsubsectionc}{0}\noindent 
        {\tenbf\thesectionc. #1}\par\vspace{5pt}}
\renewcommand{\subsection}[1] {\vspace{12pt}\addtocounter{subsectionc}{1} 
        \setcounter{subsubsectionc}{0}\noindent 
        {\bf\thesectionc.\thesubsectionc. {\kern1pt \bfit #1}}\par\vspace{5pt}}
\renewcommand{\subsubsection}[1] {\vspace{12pt}\addtocounter{subsubsectionc}{1}
        \noindent{\tenrm\thesectionc.\thesubsectionc.\thesubsubsectionc.
        {\kern1pt \tenit #1}}\par\vspace{5pt}}
\newcommand{\nonumsection}[1] {\vspace{12pt}\noindent{\tenbf #1}
        \par\vspace{5pt}}

\newcounter{appendixc}
\newcounter{subappendixc}[appendixc]
\newcounter{subsubappendixc}[subappendixc]
\renewcommand{\thesubappendixc}{\Alph{appendixc}.\arabic{subappendixc}}
\renewcommand{\thesubsubappendixc}
        {\Alph{appendixc}.\arabic{subappendixc}.\arabic{subsubappendixc}}

\renewcommand{\appendix}[1] {\vspace{12pt}
        \refstepcounter{appendixc}
        \setcounter{figure}{0}
        \setcounter{table}{0}
        \setcounter{lemma}{0}
        \setcounter{theorem}{0}
        \setcounter{corollary}{0}
        \setcounter{definition}{0}
        \setcounter{equation}{0}
        \renewcommand{\thefigure}{\Alph{appendixc}.\arabic{figure}}
        \renewcommand{\thetable}{\Alph{appendixc}.\arabic{table}}
        \renewcommand{\theappendixc}{\Alph{appendixc}}
        \renewcommand{\thelemma}{\Alph{appendixc}.\arabic{lemma}}
        \renewcommand{\thetheorem}{\Alph{appendixc}.\arabic{theorem}}
        \renewcommand{\thedefinition}{\Alph{appendixc}.\arabic{definition}}
        \renewcommand{\thecorollary}{\Alph{appendixc}.\arabic{corollary}}
        \renewcommand{\theequation}{\Alph{appendixc}.\arabic{equation}}
        \noindent{\tenbf Appendix \theappendixc #1}\par\vspace{5pt}}
\newcommand{\subappendix}[1] {\vspace{12pt}
        \refstepcounter{subappendixc}
        \noindent{\bf Appendix \thesubappendixc. {\kern1pt \bfit #1}}
        \par\vspace{5pt}}
\newcommand{\subsubappendix}[1] {\vspace{12pt}
        \refstepcounter{subsubappendixc}
        \noindent{\rm Appendix \thesubsubappendixc. {\kern1pt \tenit #1}}
        \par\vspace{5pt}}

\newcommand{\textlineskip}{\baselineskip=13pt}
\newcommand{\smalllineskip}{\baselineskip=10pt}

\def\eightcirc{
\begin{picture}(0,0)
\put(4.4,1.8){\circle{6.5}}
\end{picture}}
\def\eightcopyright{\eightcirc\kern2.7pt\hbox{\eightrm c}} 

\newcommand{\copyrightheading}[1]
        {\vspace*{-2.5cm}\smalllineskip{\flushleft
        {\footnotesize International Journal of Modern Physics C, #1}\\
        {\footnotesize $\eightcopyright$\, World Scientific Publishing
         Company}\\
         }}

\newcommand{\publisher}[2]{{\begin{center}\footnotesize\smalllineskip 
        Received #1\\
        Revised #2
        \end{center}
        }}

\def\abstracts#1#2#3{{
        \centering{\begin{minipage}{4.5in}\baselineskip=10pt\footnotesize
        \parindent=0pt #1\par 
        \parindent=15pt #2\par
        \parindent=15pt #3
        \end{minipage}}\par}} 

\newcommand{\bibit}{\nineit}
\newcommand{\bibbf}{\ninebf}
\renewenvironment{thebibliography}[1]
        {\frenchspacing
         \ninerm\baselineskip=11pt
         \begin{list}{\arabic{enumi}.}
        {\usecounter{enumi}\setlength{\parsep}{0pt}     
         \setlength{\leftmargin 12.7pt}{\rightmargin 0pt} 
         \setlength{\itemsep}{0pt} \settowidth
        {\labelwidth}{#1.}\sloppy}}{\end{list}}

\newcounter{itemlistc}
\newcounter{romanlistc}
\newcounter{alphlistc}
\newcounter{arabiclistc}

\newcommand{\fcaption}[1]{
        \refstepcounter{figure}
        \setbox\@tempboxa = \hbox{\footnotesize Fig.~\thefigure. #1}
        \ifdim \wd\@tempboxa > 5in
           {\begin{center}
        \parbox{5in}{\footnotesize\smalllineskip Fig.~\thefigure. #1}
            \end{center}}
        \else
             {\begin{center}
             {\footnotesize Fig.~\thefigure. #1}
              \end{center}}
        \fi}

\newcommand{\tcaption}[1]{
        \refstepcounter{table}
        \setbox\@tempboxa = \hbox{\footnotesize Table~\thetable. #1}
        \ifdim \wd\@tempboxa > 5in
           {\begin{center}
        \parbox{5in}{\footnotesize\smalllineskip Table~\thetable. #1}
            \end{center}}
        \else
             {\begin{center}
             {\footnotesize Table~\thetable. #1}
              \end{center}}
        \fi}

\def\@citex[#1]#2{\if@filesw\immediate\write\@auxout
        {\string\citation{#2}}\fi
\def\@citea{}\@cite{\@for\@citeb:=#2\do
        {\@citea\def\@citea{,}\@ifundefined
        {b@\@citeb}{{\bf ?}\@warning
        {Citation `\@citeb' on page \thepage \space undefined}}
        {\csname b@\@citeb\endcsname}}}{#1}}

\newif\if@cghi
\def\cite{\@cghitrue\@ifnextchar [{\@tempswatrue
        \@citex}{\@tempswafalse\@citex[]}}
\def\citelow{\@cghifalse\@ifnextchar [{\@tempswatrue
        \@citex}{\@tempswafalse\@citex[]}}
\def\@cite#1#2{{$\null^{#1}$\if@tempswa\typeout
        {IJCGA warning: optional citation argument 
        ignored: `#2'} \fi}}

\def\pmb#1{\setbox0=\hbox{#1}
        \kern-.025em\copy0\kern-\wd0
        \kern.05em\copy0\kern-\wd0
        \kern-.025em\raise.0433em\box0}

\def\fnt#1#2{\footnotetext{\kern-.3em
        {$^{\mbox{\scriptsize #1}}$}{#2}}}

\def\fpage#1{\begingroup
\voffset=.3in
\thispagestyle{empty}\begin{table}[b]\centerline{\footnotesize #1}
        \end{table}\endgroup}

\def\runninghead#1#2{\pagestyle{myheadings}
\markboth{{\protect\footnotesize\it{\quad #1}}\hfill}
{\hfill{\protect\footnotesize\it{#2\quad}}}}
\font\tenrm=cmr10
\font\tenit=cmti10 
\font\tenbf=cmbx10
\font\bfit=cmbxti10 at 10pt
\font\ninerm=cmr9
\font\nineit=cmti9
\font\ninebf=cmbx9
\font\eightrm=cmr8

\def\qed{\hbox{${\vcenter{\vbox{                        
   \hrule height 0.4pt\hbox{\vrule width 0.4pt height 6pt
   \kern5pt\vrule width 0.4pt}\hrule height 0.4pt}}}$}}

\def\bsc{{\sc a\kern-6.4pt\sc a\kern-6.4pt\sc a}}       
\def\bflatex{\bf L\kern-.30em\raise.3ex\hbox{\bsc}\kern-.14em 
T\kern-.1667em\lower.7ex\hbox{E}\kern-.125em X}

\newcommand{\be}{\begin{equation}}
\newcommand{\ee}{\end{equation}}
\newcommand{\ba}{\begin{eqnarray}}
\newcommand{\ea}{\end{eqnarray}}

\newcommand{\eps}{\epsilon}

\input amssym.def
\newsymbol\lesssim 132E
\newsymbol\gtrsim 1326 
\begin{document}
\runninghead{Long Range Correlations in Idealized Granular Flows}
{Long Range Correlations in Idealized Granular Flows}
\normalsize\textlineskip
\thispagestyle{empty}
\setcounter{page}{1}

\copyrightheading{}

\vspace*{0.88truein}

\fpage{1}
\centerline{\bf PATTERNS AND LONG RANGE CORRELATIONS}
\vspace*{0.035truein}
\centerline{\bf IN IDEALIZED GRANULAR FLOWS}
\vspace*{0.37truein}
\centerline{\footnotesize J.A.G. ORZA and R. BRITO}
\vspace*{0.015truein}
\centerline{\footnotesize\it Facultad de Ciencias F\'{\i}sicas, 
Universidad Complutense}
\baselineskip=10pt
\centerline{\footnotesize\it 28040 Madrid, Spain}
\vspace*{10pt}
\centerline{\normalsize and}
\vspace*{10pt}
\centerline{\footnotesize T.P.C. VAN NOIJE and M.H. ERNST}
\vspace*{0.015truein}
\centerline{\footnotesize\it Instituut voor Theoretische Fysica, 
Universiteit Utrecht, Postbus 80006}
\baselineskip=10pt
\centerline{\footnotesize\it 3508 TA Utrecht, The Netherlands}
\vspace*{0.225truein}
\publisher{\today}{(revised date)}
 
\vspace*{0.21truein}
\abstracts{An initially homogeneous freely evolving fluid of inelastic
hard spheres develops inhomogeneities in the flow field 
${\bf u}({\bf r},t)$ (vortices) and in the density field $n({\bf r},t)$ 
(clusters), driven by unstable fluctuations, $\delta a=\{\delta n, 
\delta{\bf u}\}$. Their spatial correlations, 
$\langle\delta a ({\bf r}, t) \delta a ({\bf r}', t)\rangle$,
as measured in molecular dynamics simulations, exhibit long
range correlations; the mean vortex diameter grows as 
$\xi(t) \propto \sqrt{\ln t}$; there occur transitions to 
macroscopic shearing states, etc. \\
The Cahn--Hilliard theory of spinodal decomposition offers a
qualitative understanding and quantitative estimates of the observed 
phenomena. When intrinsic length scales are of the order of
the system size,
effects of physical boundaries and periodic boundaries (finite size
effects in simulations) are important. 
}{}{}
\vspace*{1pt}\textlineskip 
\vspace*{1pt}\textlineskip 
\section{Introduction}
\vspace*{-0.5pt}
\noindent
Standard fluids, when out of equilibrium, will rapidly decay to 
local equilibrium within a few mean free times.
The subsequent decay of spatial inhomogeneities towards global equilibrium 
is controlled by the slow hydrodynamic time evolution.
A prototypical model for this is a system of $N$ smooth elastic hard spheres.
Compare this system to a system of 
smooth inelastic hard spheres, defined such that in the center of 
mass frame a fraction of order $\eps=1-\alpha$ (where  $\alpha$ is 
called the {\em restitution coefficient}; $0<\alpha<1$) of the 
kinetic energy is lost in each collision (for precise definitions we 
refer to Ref.\ 1).
Linear momentum is conserved during collisions. Consequently the local 
momentum density, ${\bf g}({\bf r},t)=n({\bf r},t){\bf u}({\bf r},t)$, 
or the flow velocity ${\bf u}({\bf r},t)$ is a slowly varying quantity, 
and the system can still be considered a fluid.
The lack of energy conservation makes this fluid, whether driven by 
gravity or shear stresses, or freely evolving, behave very differently 
from a standard fluid.\cite{goldhirsch} 

The system of inelastic hard spheres represents an idealized model 
for {\em rapid granular flows}, where the dynamics of individual (macroscopic)
particles is described by binary collisions, separated by free propagation 
over a typical mean free path.

To study this problem by computer simulations, we consider a 
2-dimensional system of $N$ inelastic hard disks of 
diameter $\sigma$ and unit mass,
contained in a volume $V=L^2$ with periodic boundary conditions, and use
an event-driven molecular dynamics code.\cite{allen}
At the initial time the system starts off in a spatially homogeneous 
equilibrium state of elastic hard disks ($\eps=0$).
From there on the kinetic energy is dissipated by collisions, 
which leads to cooling.

We monitor the time evolution of the system, and measure the velocity 
distribution of a particle, the time dependence of the kinetic energy, 
and the total number of collisions. These are the standard quantities 
that can be obtained from the Enskog-Boltzmann equation.

The novel feature of the present study is an analysis of the spatial 
correlations in mass and energy densities, and in flow velocities at 
different points in the system, which exhibit long range spatial correlations.
To understand such dynamic correlations one has to go beyond the 
standard Enskog-Boltzmann equation, which is based on the molecular 
chaos assumption, and use the {\em Ring}\, kinetic theory, 
which accounts for sequences of correlated binary collisions.
Ring kinetic theory and closely related mode coupling theories have been
succesful in explaining the phenomena of long time tails, and of 
long range spatial correlations in nonequilibrium stationary 
states,\cite{dorfman1,dorfman2} and in other systems that violate 
the conditions of detailed balance.\cite{harmen}
This violation also occurs in systems of inelastic hard spheres.

In fact, we have applied the Ring theory to such systems, and 
succesfully explained the behavior of the nonequilibrium pair 
distribution at late times and large distances.
Preliminary results of this investigation have been reported at 
several conferences and workshops.\cite{uswork}

The plan of this paper is as follows: section 2 describes the 
homogeneous cooling state.  Spatial fluctuations (section 3) 
drive the system away from this state through slow hydrodynamic modes 
(section 4).  Some modes are unstable and lead to phase 
separation and clustering (section 5).
Vorticity diffusion controls the long time behavior (section 6) 
and leads to scaling laws.
Section 7 deals with the effects of boundaries on the structure 
of asymptotic states.

\section{Homogeneous Cooling State}
\noindent
What are the important observations that can be deduced from 
molecular dynamics simulations, performed on a system of $N$ 
inelastic hard disks, that are prepared in a spatially 
homogeneous equilibrium state of an elastic hard disk system?
 
The system stays for {\em many} collision times in the so-called 
{\em homogeneous cooling state}, described by a single particle 
distribution function which is essentially a Maxwellian with a 
time dependent total kinetic energy $NE(t)$.\cite{goldhirsch,goldshtein,brey}
As long as the system is in this homogeneous cooling state, the 
energy per particle equals the temperature, $T(t)=E(t)$. 
This temperature decays as $t^{-2}$ for long times (see Fig.~1), 
in quantitative agreement with the predictions from kinetic 
theory, and has the form
\ba
T(t)&\equiv& \textstyle{\frac{1}{2}} v_0^2(t) = T(0) /[1+t/t_e]^2\nonumber\\
&=& T(0) \exp[-2 \gamma_0 \tau].
\label{eq:1}
\ea
Here $\gamma_0=\textstyle{\frac{1}{4}}(1-\alpha^2)=\textstyle{\frac{1}{2}}
\eps(1-\textstyle{\frac{1}{2}}\eps)$ measures the degree of inelasticity, 
and $v_0(t)$ is the thermal velocity.
The characteristic time of homogeneous cooling, $t_e=t_0/\gamma_0$, 
is proportional to the mean free time for elastic hard disks in the 
initial state, $t_0=1/\omega_0(0)=1/[\sqrt{2\pi}n\chi \sigma v_0(0)]$, 
where $\chi$ is the radial distribution function for elastic hard 
disks at contact. With this definition, the mean free path at the 
initial time equals that of the elastic hard disks 
and is given by $l_0=v_0(0)t_0$. 
The average number of collisions $\tau$, suffered by a single particle in a 
time $t$, follows by integrating $d\tau=\omega_0(t) d t$, where 
$\omega_0(t)$ is the collision frequency of inelastic hard disks, which is 
proportional to $\sqrt{T(t)}$.  The result is
\be
\gamma_0 \tau=\ln[1+t/t_e],
\label{eq:2}
\ee
where $\tau$ is related to the total number of 
collisions $C$ as $C=\textstyle{\frac{1}{2}}\tau N$.
 
\begin{figure}[htbp]
\vspace*{-0.5cm}
\begin{center}
\  \psfig{file=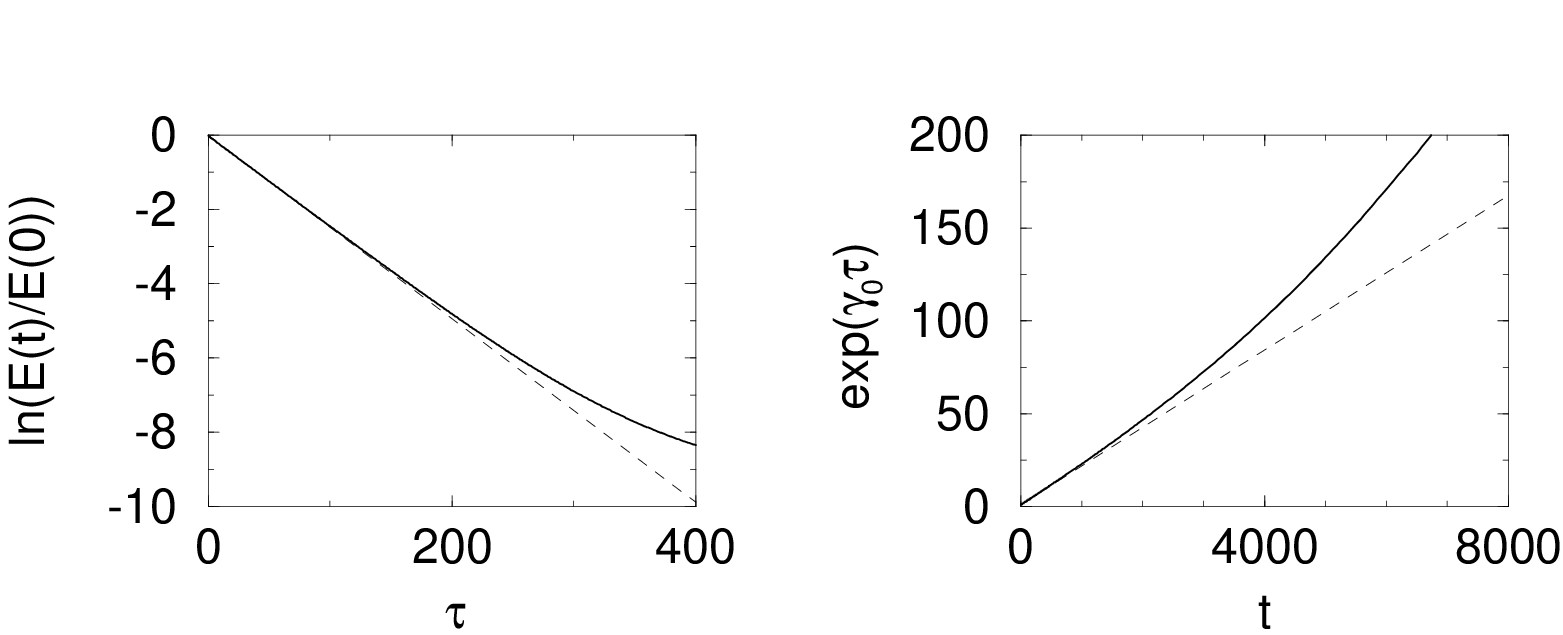,height=5cm,angle=270}\\ \
\end{center}
\fcaption{Left: Plot of the logarithm of the total energy as
a function of the number of collisions per particle $\tau$
in a simulation with $N=20000$ particles, system size of $L=253\sigma$,
and dissipation parameter   $\epsilon=1/40$.
The straight (dashed) line with slope $-2\gamma_0\simeq  -0.0247$ (Eq.~(1))
corresponds to the behavior of the homogeneous
cooling state. Deviations of $E(t)$ from the straight line around
$\tau=200-250$ are correlated with the appearance of inhomogeneities
in the density field (see behavior of $G_{nn}$ for $\tau=170-332$ in Fig.~2).
Right: Number of collisions per particle $\tau$  versus time $t$
for the same system. The straight dashed line correponds to the
homogeneous cooling prediction, Eq.~(2).}
\end{figure}

In the subsequent time evolution of the inelastic hard disk system, 
the homogeneous cooling state plays a role very similar to local equilibrium 
in ordinary fluids, and provides a conceptual basis for a hydrodynamic 
description.
However, this homogeneous cooling state does not at all behave as 
an equilibrium state of $N$ elastic hard disks with a time dependent 
temperature.
In the latter state spatial correlations are caused by hard core excluded 
volume effects and extend only over a few disk diameters $\sigma$.
In the present system of inelastic hard spheres, however, we 
have observed: ({\em i}) {\em
long range spatial} correlations extending over more than a decade of disk 
diameters; ({\em ii}) slowly growing {\em spatial 
inhomogeneities}\cite{goldhirsch,mcnamara2} in 
the local flow field ${\bf u}({\bf r},t)$ and density field $n({\bf r},t)$, 
which can be described by coupling of hydrodynamic modes, at least for
times that are not too long; ({\em iii}) 
a {\em collision number} $\tau$
that increases {\em more rapidly} than $(1/\gamma_0)\ln(1+t/t_e)$, 
and a {\em cooling law} that decays {\em more 
slowly}\cite{goldhirsch,mcnamara2} than the prediction of 
Eq.\ (\ref{eq:1}), which is illustrated in Fig.\ 1.

\section{Spatial Correlations\label{sec:3}}
\noindent
To obtain a better understanding of the dynamics of a freely 
evolving system of these inelastic hard disks, we follow the local 
mass density $n({\bf r},t)$, the local momentum density 
${\bf g}({\bf r},t)=n({\bf r},t) {\bf u}({\bf r},t)$ and the 
local energy per particle $e({\bf r},t)$ in a single realization 
of a system of $N=20000$ particles by means of computer simulations,
where the inelasticity parameter is $\eps=1/40$ and the volume 
fraction $\phi=\textstyle{\frac{1}{4}}\pi n\sigma^2 \simeq 0.245$, 
corresponding to a system size $L\simeq 253 \sigma$.
 
Visual observation of snapshots of $n({\bf r },t)$- and 
${\bf g}({\bf r},t)$-fields  show that at early times 
($\tau=82$), the momentum density does not exhibit much 
visible structure, and the mass density is totally homogeneous.
At $\tau=120-170$ the mass density is still homogeneous, 
but the flow field starts to show vortices after coarse graining over cells
of $5\sigma\times 5 \sigma$.
At the latest time ($\tau=332$) the flow field shows a 
pronounced structure of vortices, also in the 
fine grained ${\bf u}({\bf r},t)$-field,  with a typical diameter $\xi$ of 
the order of $\textstyle{\frac{1}{4}}$ of the system size $L$, 
and the mass density starts to show barely visible inhomogeneities in the 
fine grained $n({\bf r},t)$-field, but clearly visible in the coarse 
grained $n({\bf r},t)$-fields.
 
To quantify these observations we have measured the spatial correlation 
functions of the microscopic densities, i.e.
\be
n^2 G_{ab}({\bf r},t)=\frac{1}{V}\int d{\bf R} \langle \delta 
\hat{a}({\bf R}+{\bf r},t)\delta \hat{b} ({\bf R},t)\rangle.
\ee
Here the labels $a$, $b$ take the values $\{n,e,\alpha,\parallel,\perp\}$, 
and refer respectively to the microscopic mass density, momentum density 
and local energy per particle,
\ba
\delta \hat{n}({\bf r},t) &=&\sum_i \delta({\bf r}_i(t)-{\bf r})-n\nonumber\\
n \hat{u}_\alpha({\bf r},t)&=&\sum_i v_{i\alpha}(t) 
\delta({\bf r}_i(t)-{\bf r})\nonumber\\
n\delta \hat{e}({\bf r},t)&=&\sum_i \textstyle{\frac{1}{2}}
\left(v_i^2(t)-v_0^2(t)\right)\delta({\bf r}_i(t)-{\bf r}).
\ea
The carets denote microscopic quantities.
Moreover, $v_0(t)=\sqrt{2T(t)}$ is the thermal velocity and ${\bf v}_i$, 
${\bf r}_i$ are the velocity and position of the $i$-th particle.
Greek labels $\alpha,\beta=\{x,y\}$ refer to Cartesian components.
The second rank tensor field $G_{\alpha\beta}({\bf r},t)$ has transverse 
$(\perp)$ and longitudinal $(\parallel)$ components
\ba
G_\parallel(r,t)&=&\hat{\bf r}_\alpha\hat{\bf r}_\beta 
G_{\alpha \beta}({\bf r},t)\nonumber\\
G_\perp(r,t)&=&\hat{\bf r}_{\perp\alpha}\hat{\bf r}_{\perp\beta} 
G_{\alpha \beta}({\bf r},t),
\ea
where $\hat{\bf r},\,\,\hat{\bf r}_{\perp}$ are unit vectors, parallel and 
perpendicular to the relative position ${\bf r}$.
As long as the system remains spatially homogeneous with a 
vanishing flow velocity, the local energy and temperature 
fluctuations are the same.
However, as soon as the average flow velocity ${\bf u}({\bf r},t)$ 
is nonvanishing, the granular temperature is the average energy 
per particle in the local rest frame.
 
\begin{figure}[htbp]
\vspace*{-0cm}
\begin{center}
\  \psfig{file=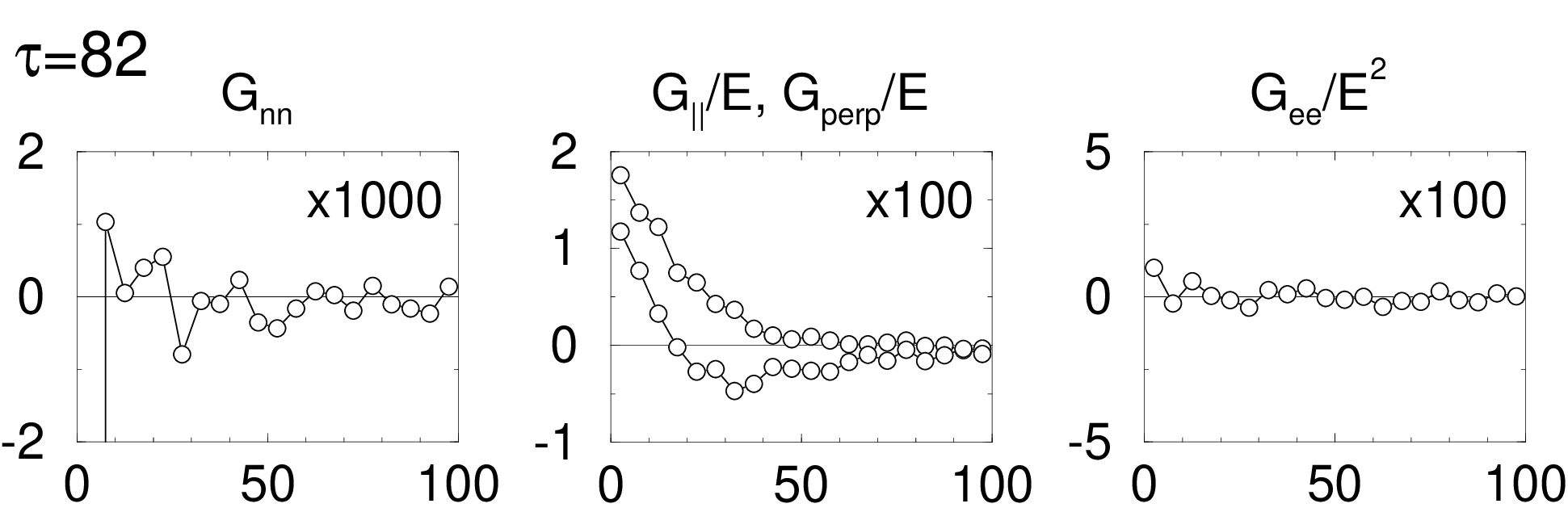,height=4cm,angle=270}\\[2mm]
\  \psfig{file=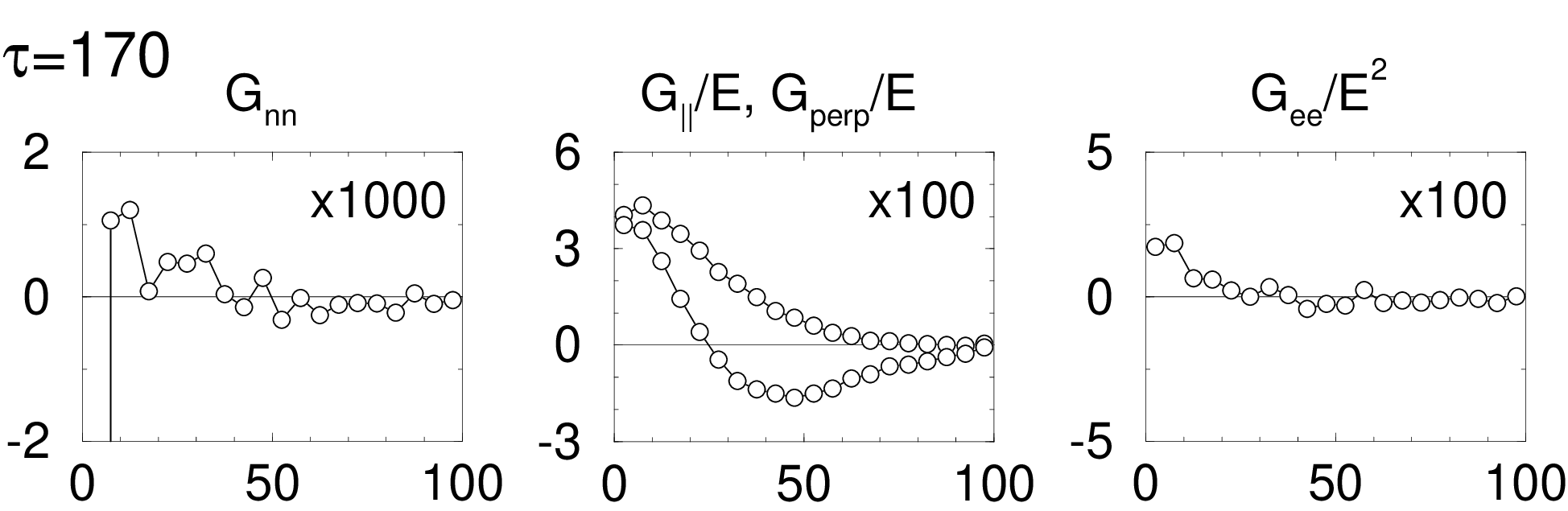,height=4cm,angle=270}\\[2mm]
\  \psfig{file=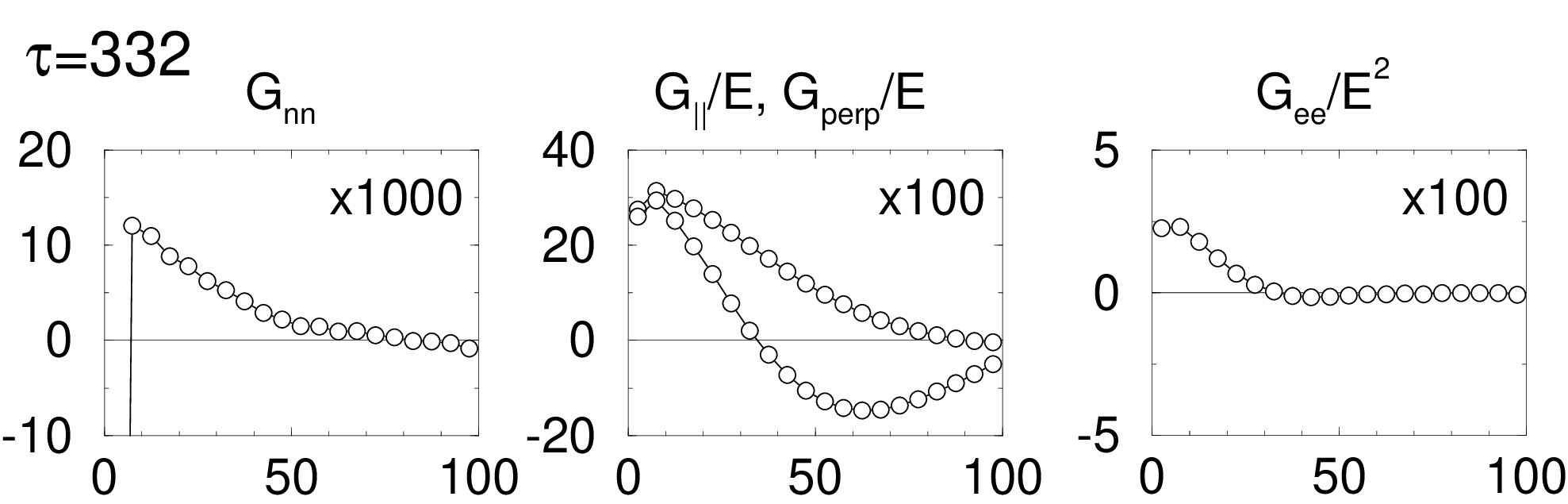,height=4cm,angle=270}\\[5mm] \ 
\end{center}
\fcaption{Spatial correlations of relative fluctuations around the 
homogeneous cooling state $G_{ab}({\bf r},t)/E^p$ in mass densities 
($nn$), flow velocities ($\parallel$, $\perp$) and energy 
densities ($ee$) 
($p=0,1,2$ respectively), plotted versus $r/\sigma$ 
at a fixed number of collisions $\tau=2C/N$ per particle, measured in 
a single run on a system with $N=20000$ particles in a volume $V=L^2$ 
with $L=253 \sigma$ and dissipation parameter $\eps=1/40$. Notice the 
vertical scales, showing magnitudes and growth rate 
of the fluctuations: at $\tau=332$, $G_\parallel$ and $G_\perp$
have increased (20$\times$), $G_{nn}$ (10$\times$) and $G_{ee}$ (1$\times$).
The measurements of $G_{ab}({\bf r},t)$, shown here, are coarse grained
by collecting data points in bins of width $\Delta r=5\sigma$. For 
bin sizes of $\Delta r =0.2\sigma$ there is already a 
`signal over noise level'
in $G_\parallel$ and $G_\perp$ at $\tau=82$, whereas the signals in 
$G_{nn}$  and $G_{ee}$ just start to rise above noise level at short
distances $r<30\sigma$ at $\tau\simeq 220$. }
\label{fig:1}
\end{figure}
The measurement of correlation functions in a single realization is shown 
in Fig.\ 2.
The results are remarkable!
Even at the earliest time ($\tau=82$), long before 
any vortices are visible, long range spatial correlations have developed 
between the local flow velocities at different points in the fluid with 
a typical correlation length $\xi\simeq 30\sigma$ 
($\xi$ is the location of the minimum in $G_\perp(r,t)$), which 
gradually increases to about $60\sigma$ at $\tau=332$.
The negative minimum in $G_\perp(r,t)$ clearly signals the 
presence of vortices with a diameter $\xi$.  It is also interesting to 
compare the noise level and magnitude (vertical axes in Fig.\ 2) of all 
fluctuations, relative to the homogeneous cooling state, and correlate the 
signals above noise level and their growth rates with the visual observations 
of the $n({\bf r},t)$- and ${\bf g}({\bf r},t)$-snapshots.
 
The dynamics of the phenomena observed seems to be controlled for a long
time by the transverse flow field or vorticity modes.
We {\em conjecture} that there exists an extended time regime in 
which these modes describe the behavior of the spatial correlation 
functions, at least of those correlation functions that 
involve the flow fields.
We shall test this conjecture by analyzing the simulation results.
 
Moreover, we note that in normal fluids with elastic collisions, times 
$\tau\gtrsim 5$ are already considered long in kinetic theory.
The long time tail in the velocity autocorrelation function of elastic
hard spheres is typically observed in the ranges $10\lesssim\tau\lesssim
60$.  The range of the spatial correlation functions $G_\perp(r,t)$ 
and $G_\parallel(r,t)$, observed here, extends far beyond any 
{\em static} correlation length (a few $\sigma$'s) and far beyond the 
mean free path ($l_0\simeq 0.8\sigma$ at this density).
 
\section{Hydrodynamic Dispersion Relations\label{sec:4}}
\noindent
The previous observations on large spatial and temporal 
scales suggest that the long wavelength components of the 
microscopic fluctuations are described by {\em linearized} 
(because we are dealing with fluctuations) {\em hydrodynamics}.
This should hold as long as fluctuations have not grown to 
macroscopic size.
Beyond this time the evolution is controlled by the full 
nonlinear hydrodynamic equations.
 
The fluctuations discussed in the previous sections are taken 
{\em relative} to the homogeneous cooling state.
Consequently, we linearize the hydrodynamic equations around 
this state,\cite{goldhirsch,mcnamara1} and consider the 
Fourier components of the hydrodynamic fields $\delta n({\bf k},t)$, 
$u_\alpha ({\bf k},t)/v_0(t)$ and $\delta e({\bf k},t)/v_0^2(t)$ 
where $T(t)=\textstyle{\frac{1}{2}}v_0^2(t)$ is the temperature 
in the homogeneous cooling state.
The resulting set of linear equations contains still time 
dependent coefficients, such as the local pressure, 
proportional to $T(t)$, and transport coefficients, which are 
assumed to be independent of the dissipation parameter $\eps$, 
and proportional to the average collision frequency 
$\omega_0(t)\propto \sqrt{T(t)}$.
Hence, the kinematic viscosity of inelastic hard disks is 
$\nu(t)=\nu \exp(-\gamma_0\tau)$, where $\nu=\eta/m n$ is 
the corresponding viscosity of elastic hard disks at the initial time.
 
By changing to the time variable $\tau$, introduced in 
Eq.\ (\ref{eq:2}), the linearized hydrodynamic equations transform 
into a set of coupled linear equations with constant coefficients.
The resulting eigenmodes have the form $\psi_\mu({\bf k},\tau)=
\psi_\mu({\bf k}) \exp[z_\mu(k)\tau]$, where 
$\mu=\{\perp,H,\sigma=\pm\}$ labels the shear mode ($\perp$), 
the heat mode ($H$) and the sound modes ($\sigma=\pm$).
Note that the modes only show exponential behavior if time 
is measured by counting collisions.
In real time the modes decay algebraically.
The shear mode $\psi_\perp({\bf k},\tau)$ is 
$u_\perp({\bf k},t)/v_0(t)$, and the heat and sound 
modes are linear combinations of the relative fluctuations 
$\delta n({\bf k},t)$, $u_\parallel({\bf k},t)/v_0(t)$ and 
$\delta e({\bf k},t)/v_0^2(t)$.
Here the subscripts longitudinal ($\parallel$) and transverse 
($\perp$) refer to the direction of ${\bf k}$.
 
The dispersion relations for the eigenvalues $z_\mu$ of heat and 
shear mode are shown in Fig.\ 3 as a function of the rescaled 
wavenumber $q=k\sigma/\sqrt{\gamma_0}$.
\begin{figure}[htbp]
\begin{center}
\ \psfig{file=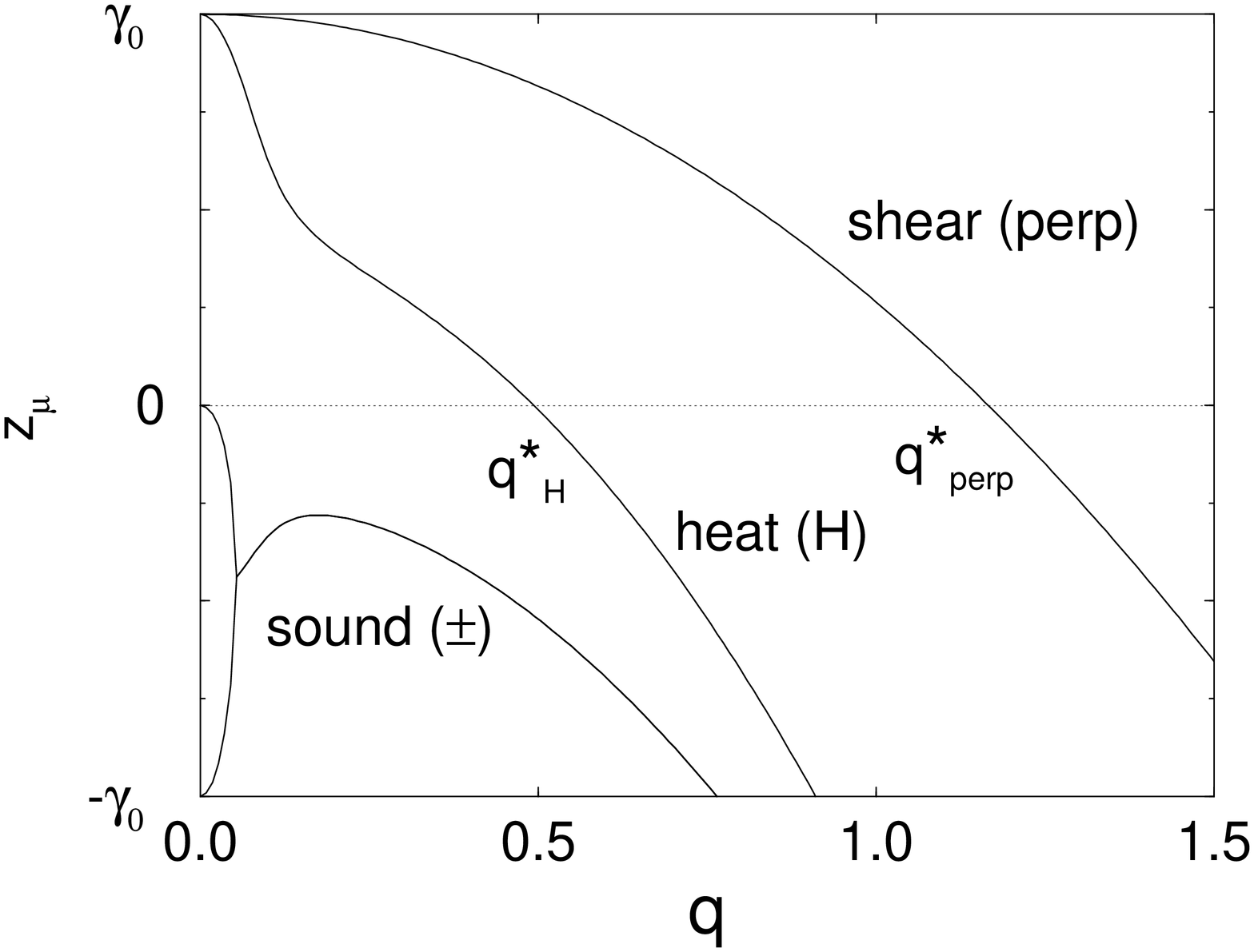,height=6cm,angle=270} \ \\ \  
\end{center}  
\fcaption{Dispersion relations for the decay rates $z_\mu$ of 
hydrodynamic modes ($\mu=\{\perp,H,\pm\}$) in units $\gamma_0$, 
plotted versus the reduced wavenumber $q=k\sigma/\sqrt{\gamma_0}$ to 
make $z_\perp/\gamma_0$ independent of dissipation parameter $\gamma_0=
\textstyle{\frac{1}{2}}\eps(1-\textstyle{\frac{1}{2}}\eps)$. Numerical 
values are calculated from the Enskog-Boltzmann equation for elastic hard 
disks at volume fraction $\phi=\textstyle{\frac{1}{4}}\pi n\sigma^2\simeq 
0.245$. Wavenumber $q_H^\ast$ and $q_\perp^\ast$ are the thresholds of 
stability for heat and shear modes.}
\label{fig:2}
\end{figure}
These functions intersect the $q$-axis respectively at 
$q_H^\ast$ and $q_\perp^\ast$, where $q_\mu^\ast=k_\mu^\ast
\sigma/\sqrt{\gamma_0}$.
At volume fraction $\phi\simeq 0.245$ one has 
$q_H^\ast\simeq 0.50$ and $q_\perp^\ast\simeq1.17$.
The numerical values of the transport coefficients are calculated from 
the Enskog theory for elastic hard disks\cite{chapman} at a volume 
fraction $\phi\simeq 0.245$ and initial temperature $T(0)=1$, 
used in the present simulations.
We only quote the result for the shear mode or transverse velocity 
field,\cite{goldhirsch,brey,mcnamara1}  which decays as
\ba
u_\perp({\bf k},t)&\sim& v_0(t)\exp[z_\perp(k)\tau]\nonumber\\
&=&v_0(0) (1+t/t_e)^{-\nu k^2 t_e},
\ea
where the dispersion relation for the relaxation rate is:
\be
z_\perp(k)=\gamma_0(1-\nu k^2 t_e)=\gamma_0 (1-\hat{\nu} q^2),
\ee
with  $\hat{\nu}=\nu t_0/\sigma^2$.
If the viscosity of the inelastic hard disks depend 
on the dissipation parameter $\eps$ only through 
$\sqrt{T(t)}$---as we have assumed here---, then $z_\perp(k)/\gamma_0$ 
is a universal function of $q=k\sigma /\sqrt{\gamma_0}$, 
independent of the dissipation parameter $\gamma_0$.

\section{Instabilities and Clustering}
\noindent
The spectrum of eigenvalues in Fig.\ 3 shows that the relative 
fluctuations with respect to the homogeneous cooling state contain several 
{\em unstable} modes with an {\em exponential} growth rate, 
$\exp[z_\mu(k)\tau]$, if time is measured in collision numbers $\tau$.
These unstable modes, $\psi_\mu({\bf k},\tau)$, are the long wavelength
shear mode ($\mu=\perp$) with $k<k_\perp^\ast$, and the long wavelength
heat mode $\psi_H(k,\tau)$ with $k<k_H^\ast<k_\perp^\ast$.
The sound modes are stable for all wavelengths.
The dynamics of these slow modes determines the time 
evolution of the long wavelength fluctuations, away from the 
homogeneous cooling state, and
determines a mechanism for initial pattern selection.
In the initial stages spontaneous fluctuations with respect to the 
homogeneous cooling state occur on all possible wavelengths.
Those with $k$ smaller than $k_H^\ast$ or $k_\perp^\ast$ will grow at an
exponential rate $\exp[z_\mu(k)\tau]$.
 
The fastest growing mode is the one with the smallest wavenumber, 
as can be seen from the dispersion relations in Fig.\ 3.
In finite systems with periodic boundary conditions, as 
used in the present computer simulations, the smallest wavenumber 
allowed is $k_m=2\pi/L$.
As long as $k_m<k_H^\ast$, unstable shear modes as well as heat 
modes can be excited, and we can estimate the {\em onset times} 
$\tau_\mu$ for the shear and heat mode instabilities by the criterion
\be
z_\mu(k_m) \tau_\mu \simeq 1\qquad (\mu=\perp,H),
\label{eq:crit}
\ee
which implies that the amplitude of relative fluctuation has 
increased by a factor $e$ above its normal level.
As long as the fluctuations are small and, hence, can be described by 
{\em linearized hydrodynamic} equations, the vorticity mode does 
not couple to density and energy fluctuations, and the onset for 
the appearance of vortex structure is the time $\tau_\perp$, 
as defined in Eq.\ (\ref{eq:crit}).
The fluctuations in mass and energy densities, however, do couple 
to the unstable heat mode, and the onset time for observable 
inhomogeneities in the mass density (clustering) and energy field will 
be of order $\tau_H$.
 
For the system used in the simulations of section 3 
with $L=253 \sigma$ and $\eps=0.025$ one has $q_m\simeq 0.22$.
The corresponding rate constants $z_\mu(k_m)$ can be read off 
from Fig.\ 3 and yield estimates of the onset times for the 
appearance of vorticity ($\tau_\perp\simeq 83$) and for 
clustering in the density ($\tau_H\simeq 225$).
Inspection of the configurations $n({\bf r},t)$ and 
${\bf g}({\bf r},t)$, as well as the correlations in Fig.\ 2, shows that the 
structures in $G_{nn}(r,t)$ and $G_{ee}(r,t)$ are barely 
noticeable at $\tau=170$, and well developed at $\tau=332$, 
consistent with the theoretical estimates for $\tau_\perp$ and $\tau_H$.
 
Suppose we decrease the system size or decrease $\eps$, such 
that $q_m=2\pi\sigma/[L\sqrt{\gamma_0}]$ increases and moves 
into the interval $(q_H^\ast,q_\perp^\ast)$.
Then only shear modes can be excited and the onset time $\tau_\perp$ for
the appearance of vortex structures will increase like $1/z_\perp(k_m)$ 
with increasing $q_m$.
The vorticity will grow, but density and energy fields will remain 
{\em spatially homogeneous}, as long as linearized hydrodynamics is adequate.
An example of such a system is realized at $N=5000$, $L=126 \sigma$ 
and $\eps=0.007$, where $q_m\simeq 0.84$, which is located
in the middle of the interval $(q_H^\ast,q_\perp^\ast)$.
The simulations show that the total energy follows the {\em homogeneous 
cooling law}, $E(t)=E(0)\exp[-2\gamma_0\tau]$, up to $\tau\simeq 1100$.
Of course, in all cases where $k_m<k_\perp^\ast$, eventually the 
fluctuations will grow to macroscopic size, and after a very long time 
nonlinear hydrodynamics takes over,\cite{goldhirsch} and leads again 
to inhomogeneities in density- and energy fields.
 
If one still further increases $q_m$, it will pass the stability 
threshold $q_\perp^\ast$, and none of the unstable modes is excitable.
All hydrodynamic modes become stable, and the system will remain in 
the homogeneous cooling state for {\em all times}. McNamara and 
Young\cite{mcnamara2} call this state the kinetic state. An example is 
the system $\{N=5000,L=126 \sigma,\eps=0.001\}$, 
where $q_m\simeq 2.2 >q_\perp^\ast$.
Even after $\tau=2044$ no inhomogeneities in density or flow fields have
developed, and the homogeneous cooling law is still obeyed, as 
can be inferred from the measured correlation functions 
$G_{ab}({\bf r},t)$ and the measured kinetic energy $E(t)$.

\section{Vorticity Diffusion and Scaling\label{sec:6}}
A rough estimate of the behavior of the pair correlation function can 
be obtained from the Cahn-Hilliard theory\cite{cahn} 
for the dynamics of phase separation.
In this theory one calculates the {\em structure factors}, defined 
as the Fourier transforms $S_{ab}({\bf k},t)=V^{-1}\langle 
\delta \hat{a}({\bf k},t)\delta \hat{b}({\bf k},t)\rangle$ of 
the correlation functions $G_{ab}({\bf r},t)$, from the long 
wavelength behavior of the unstable hydrodynamic (macroscopic) modes 
$\delta a({\bf k},t)$. 
Very recently Deltour and Barrat used this theory to calculate the 
structure factor $S_{nn}$ for a fluid of inelastic hard disks.\cite{deltour}

Here a similar estimate for the structure factor 
$S_\perp(k,t)=V^{-1}\langle|\hat{u}_\perp({\bf k},t)|^2\rangle$ of 
the vorticity field yields $S_\perp(k,t)\propto v_0^2(t) 
\exp[2 z_\perp(k)\tau]\propto\exp[-2\nu t_0 k^2 \tau]$.
On the basis of the conjecture, formulated in section 3, 
the contributions of heat and sound modes can be neglected.
The correlation functions $G_\perp(r,t)$ and $G_\parallel(r,t)$ 
can then be obtained by taking inverse Fourier transforms of 
$(\hat{\bf k}\cdot\hat{\bf r})^2 S_\perp(k,t)$ and 
$[1-(\hat{\bf k}\cdot\hat{\bf r})^2] S_\perp(k,t)$ respectively, 
where $\hat{\bf a}={\bf a}/|{\bf a}|$.
They have the generic scaling form $\tau^{-1}F_\mu(r/\sqrt{\nu t_0\tau})$ 
with $\mu=\{\perp,\parallel\}$.
The explicit form of $F_\mu(x)$ is not used in the arguments
presented below.

We recall that the average diameter $\xi(t)$ of a vortex can be 
identified with the location of the minimum of $G_\perp(r,t)$.
It then follows from the scaling form and Eq.\ (\ref{eq:2}) that 
\be
\xi(t)\simeq a_0 \sqrt{\nu t_0\tau} \propto \sqrt{\nu t_e\ln(t/t_e)},
\ee
where the constant of proportionality $a_0$ is independent of the 
volume fraction $\phi$ and of the dissipation parameter $\eps$.
 For {\em large systems} the value of $a_0$, measured in the 
simulations, is typically $a_0\simeq 4.0-5.3$ or $3.8-4.0$ in 
the $\tau$-intervals ($10-80$) or ($82-216$) respectively for 
systems with parameters $\{N=5\times 10^4,\eps=0.1,\phi=0.4\}$ 
or $\{N=2\times 10^4,\eps=0.025,\phi=0.245\}$.
For small systems (typically $N=5000$, $\eps<0.1$), $\xi(t)$ 
saturates to a constant value and diffusive growth is absent due 
to interference effects caused by the periodic boundaries.

The diffusive growth of the average vortex diameter offers strong 
support for the validity of our conjecture that the long time dynamics 
on large spatial scales is mainly governed by vorticity diffusion.
More refined Ring kinetic theory or mode coupling theories provide 
detailed information\cite{us} about the explicit analytic 
forms of $G_\mu(r,t)$.
It is found that data collapse for $G_\mu(r,t)$ with 
$\mu=\{\perp,\parallel\}$ 
{\em only} hold for large distances, where $r >\sqrt{\nu t_0 \tau}$, and 
large times, where $\tau>\tau_e= (\ln{2})/\gamma_0$. 
Here $\tau_e\equiv\tau(t_e)$ is the characteristic time scale for 
homogeneous cooling, as defined through Eq.\ (\ref{eq:2}).

We conclude that our theoretical analysis has provided a consistent picture of 
the long range dynamic correlations that have built up through the dissipative 
dynamics of inelastic hard sphere systems.
This dynamics violates the conditions of detailed balance.

The theoretical analysis also predicts the existence of {\em algebraic} long 
range correlations $\propto 1/r^d$, where $d$ is the dimensionality of 
the hard sphere system.
Similar algebraic tails have been observed in lattice gas automata with 
collision rules that violate the conditions of detailed balance.\cite{harmen}
The complete analysis and theoretical predictions, which have no adjustable 
parameters, as well as extensive comparisons with the results of computer 
simulations will be reported elsewhere.\cite{us}

A comparison with the simulation results for a single realization is shown in 
Fig.~4.
\begin{figure}[htbp]
\begin{center}
\  \psfig{file=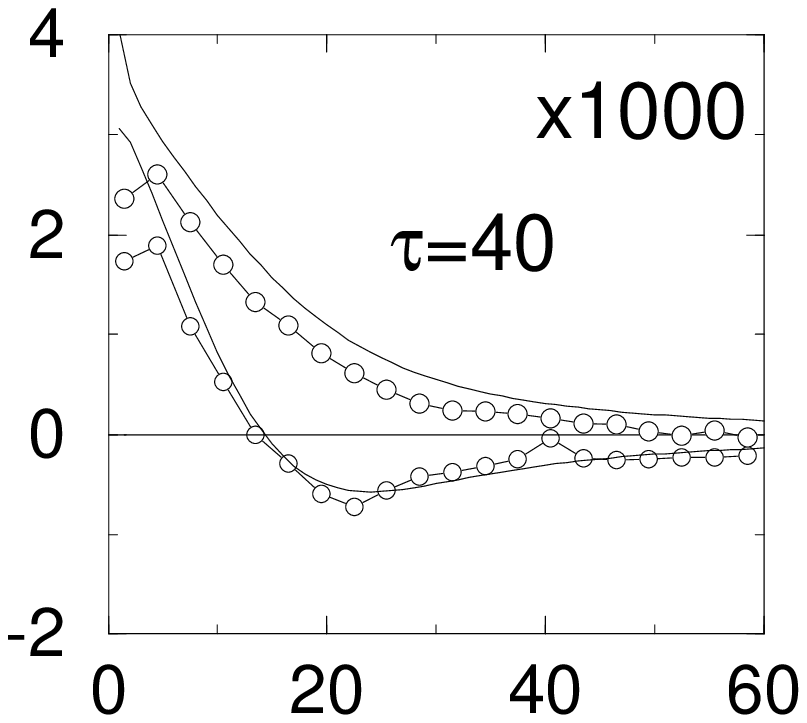,height=4.5cm,angle=270}\\
\end{center}
\label{fig:3}
\fcaption{Comparison of theory (solid line) and simulations (circles) of 
$G_\parallel(r,t)$ and $G_\perp(r,t)$ as functions of $r/\sigma$ at 
$\tau=2C/N\simeq 40$ collisions per particle in a system with $N=5000$, 
$L=126\sigma$, dissipation parameter $\eps=0.1$, homogeneous cooling 
length $l_e\simeq 7\sigma$, with unstable heat and shear modes 
($q_m=2\pi\sigma/[L\sqrt{\gamma_0}]\simeq 0.23 < q_H^\ast$), and 
estimated transition to the shearing state at $\tau_{\rm tr}=
L^2/[64\nu t_0] \simeq 325$.}
\end{figure}

There is a quantitative agreement between theory and simulations for all 
$r$-values, provided $\tau$ is sufficiently small 
($\tau\lesssim 60$ in Fig.\ 4) to avoid effects from nonlinear 
hydrodynamics, and provided system size 
$L$ and/or dissipation parameter $\eps$ are sufficiently large to avoid 
interference effects from periodic boundaries.

\newpage

\section{Transition to Shearing States/Boundary Effects\label{sec:7}}
What are the asymptotic states observed in computer simulations?
In {\em large} but finite systems with $L$ and $\eps$ large 
enough (that is $q_m<q_H^\ast$), 
so that all unstable modes can be excited, the mean vortex 
diameter can not grow indefinitely.
It will typically stop growing at a {\em transition time} $\tau_{\rm tr}$, 
where $\xi (\tau_{\rm tr})\simeq \textstyle{\frac{1}{2}}L$, and the 
periodic boundaries {\em force} a transition to an inhomogeneous shearing state.
The dynamics of this transition, caused by fusion of vortices of equal sign 
into shearing layers, is well illustrated in the snapshot of the 
flow field in Fig.\ 10b of Ref.\ 10.
In this case, the transition time $\tau_{\rm tr}$ can be estimated 
from the growth law $\xi(\tau)\simeq4\sqrt{\nu t_0\tau}$,
and yields $\tau_{\rm tr}\simeq L^2/64 \nu t_0$, which is proportional
to the volume of the system, and independent of the dissipation
parameter $\eps$.  For the system with $\{N=5000,L=126,\eps=0.1\}$ the
theoretical estimate gives $\tau_{\rm tr}\simeq 325$, whereas the simulations
yield $\tau_{\rm tr}\simeq 400\pm 25$.
In the system of Ref.\ 2 with $\{N=4\times 10^4,
L\simeq 793, \eps=0.4\}$ the above estimate gives $\tau_{\rm tr}\simeq 525$,
whereas the value reported for the simulation is
$\tau_{\rm tr}=2 C/N\simeq 600$.

However, for systems with $\{N=5000,
L\simeq 126, \epsilon\lesssim 0.01\}$, and 
for the small systems of Ref. 10 with  $\{N=1024, 
L=57, \eps\simeq 0.18-0.45\}$ {\em diffusive growth} of the mean 
vortex diameter is {\em absent} due to interference effects resulting from 
the periodic boundaries, as the stability wavelength for the 
heat mode $l_H=2\pi/k_H^*$ is of order $L$ (see discussion below).
As diffusive growth is absent, the transition time to the shearing
state,
$\tau_{tr}=2C/N\simeq 1400-1600$, observed in the simulations of 
Fig.~5, and of Ref.~10, has no connection to the theoretical estimate 
$\tau_{tr}=L^2/64\nu t_0$, valid for {\em large systems}.

\begin{figure}[htbp]
\begin{center}
\  \psfig{file=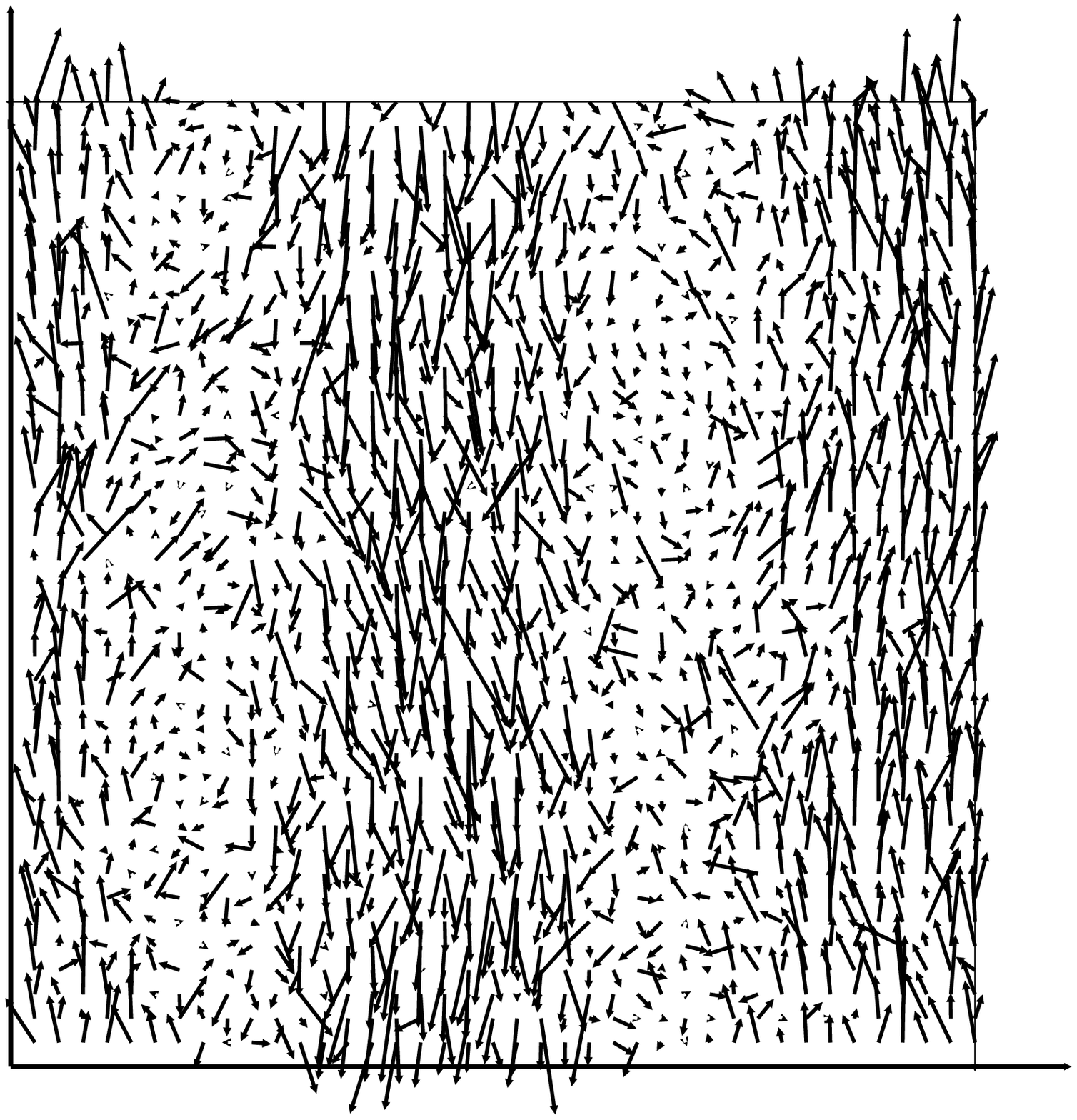,height=6cm,angle=270}\\
\end{center}
\label{fig:4}
\fcaption{Flow field, coarse grained over cells of $3\sigma\times 3\sigma$,
showing a transition at $\tau_{\rm tr}=2C/N\simeq 1600$ collisions per
particle to a shearing state with a homogeneous density distribution in a
small system with $N=5000$, $L=126\sigma$, dissipation parameter
$\eps=0.01$, stability wavelength $l_H\simeq126\sigma$, and
homogeneous cooling length $l_e\simeq 160\sigma$ (see section 7).
}
\end{figure}

In the inelastic hard sphere system under discussion, there exists
a number of intrinsic dynamical length scales: the mean free path 
$l_0$, the homogeneous cooling length
$l_e=v_0(0)t_e=l_0/{\gamma_0}$; the mean vortex size $\xi(t)$; the
stability wavelength, $l_\mu=2\pi/k_\mu^*$, for the 
stability of the heat ($\mu=H$) and shear mode ($\mu=\perp$).

If any of these length scales become of the order of the 
system size $L$, the effects of the boundaries dominate the
behavior of the system. If one is dealing with finite geometries,
the effects of physical boundaries become important, and one needs to 
model particle-wall interactions.

In case one is dealing with computer simulations, in which periodic
boundary conditions are used, the intrinsic length scales should be less
than $\textstyle{\frac{1}{3}}L$, say, in order to minimize finite size
effects. If, however, one of these length scales becomes of order $L$, 
the long time dynamics of the system is totally controlled by the 
artificial periodic boundary conditions, and the final states are
artifacts of these unphysical boundaries. 

The above discussions show that the homogeneous and inhomogeneous 
shearing states, as well as the kinetic state, discussed in the 
previous sections, and reported in the 
literature,\cite{goldhirsch,mcnamara2,deltour}
do not represent physical properties of macroscopic (idealized) granular 
flows, but are caused by the artificial periodic boundary conditions.
It is also interesting to observe that in cases where the boundary 
conditions have been modelled in a physically more realistic way and in 
different geometries\cite{esipov}---i.e.\ by a wall kept at a constant 
temperature---different asymptotic states have been observed.
This discussion suggests that more realistic modelling of collisions 
between granular particles and physical walls is imperative for 
understanding the asymptotic states of granular flows contained 
in finite geometries.

\nonumsection{Acknowledgements}
\noindent
Two of us (R.B. \& T.v.N.) acknowledge support of the 
foundation `Fundamenteel Onderzoek der Materie (FOM)', which is 
financially supported by the Dutch National Science Foundation (NWO).
One of us (R.B.) acknowledges further financial support from  
Universidad Complutense through the program 
`Estancias Breves en el Extranjero' and DGICYT (Spain) number PB94-0265.
One of us (M.H.E.) acknowledges support from the Offices of International 
Relations of Universidad Complutense and Universiteit Utrecht for his 
stay in Madrid.

\nonumsection{References}

\end{document}